\documentclass[%
reprint,
superscriptaddress,
preprintnumbers,
amsmath,
amssymb,
aps,
floatfix,
]{revtex4-2}
\usepackage{graphicx}
\usepackage{dcolumn}
\usepackage{bm}
\usepackage[utf8]{inputenc}
\usepackage[T1]{fontenc}
\usepackage{mathptmx}
\usepackage{textcomp}
\usepackage{siunitx}
\usepackage{bigints}

\usepackage{hyperref}
\usepackage[switch]{lineno}
\usepackage[table,xcdraw,dvipsnames]{xcolor}
\usepackage{booktabs}
\usepackage{float}
\usepackage{color}
\usepackage{ulem}
\usepackage{multirow}

\begin{document}

\title[Signatures of higher-order skyrmionic textures revealed by magnetic force microscopy]{Signatures of higher order skyrmionic textures revealed by magnetic force microscopy}

\author{Sabri~Koraltan}
\email{sabri.koraltan@tuwien.ac.at}
\affiliation{Institute of Applied Physics, TU Wien, Vienna, A-1040, Austria}
\affiliation{Physics of Functional Materials, Faculty of Physics, University of Vienna, A-1090 Vienna, Austria}%
\affiliation{Research Platform MMM Mathematics-Magnetism-Materials, University of Vienna, A-1090 Vienna, Austria.}%
\affiliation{Vienna Doctoral School in Physics, University of Vienna, A-1090 Vienna, Austria}%
\affiliation{Institute of Physics, University of Augsburg, D-86135 Augsburg, Germany}
\author{Joe~Sunny}
\affiliation{Institute of Physics, University of Augsburg, D-86135 Augsburg, Germany}
\author{Tamer~Karaman}
\affiliation{Institute of Physics, University of Augsburg, D-86135 Augsburg, Germany}
\author{Reshma~Peremadathil-Pradeep}
\affiliation{Magnetic \& Functional Thin Films Laboratory, Empa, Swiss Federal Laboratories for Materials Science and Technology, Ueberlandstrasse 129, 8600 Dübendorf, Switzerland}
\author{Emily~Darwin}
\affiliation{Magnetic \& Functional Thin Films Laboratory, Empa, Swiss Federal Laboratories for Materials Science and Technology, Ueberlandstrasse 129, 8600 Dübendorf, Switzerland}
\author{Felix~Büttner}
\affiliation{Institute of Physics, University of Augsburg, D-86135 Augsburg, Germany}
\affiliation{Helmholtz-Zentrum Berlin, 14109 Berlin, Germany}
\author{Dieter~Suess}
\affiliation{Physics of Functional Materials, Faculty of Physics, University of Vienna, A-1090 Vienna, Austria}%
\affiliation{Research Platform MMM Mathematics-Magnetism-Materials, University of Vienna, A-1090 Vienna, Austria.}%
\author{Hans~Josef~Hug}
\affiliation{Magnetic \& Functional Thin Films Laboratory, Empa, Swiss Federal Laboratories for Materials Science and Technology, Ueberlandstrasse 129, 8600 Dübendorf, Switzerland}
\author{Manfred~Albrecht}
\affiliation{Institute of Physics, University of Augsburg, D-86135 Augsburg, Germany}

\date{\today}

\begin{abstract}
Higher-order skyrmions and antiskyrmions are topologically protected spin textures with an integer topological charge other than $\pm 1$ and nucleate from topological point defects in regular Bloch walls, known as vertical Bloch lines. So far, they have only been observed using Lorentz transmission electron microscopy. In this work, we show that higher-order spin textures coexisting in Co/Ni multilayers at room temperature can be visualized by high-resolution magnetic force microscopy (MFM). The experimental results are supported by micromagnetic simulations confirming that different spin objects give rise to distinct MFM contrast in full agreement to our observations.
\end{abstract}

\maketitle

\begin{figure*}
  \centering
  \includegraphics[width=\textwidth]{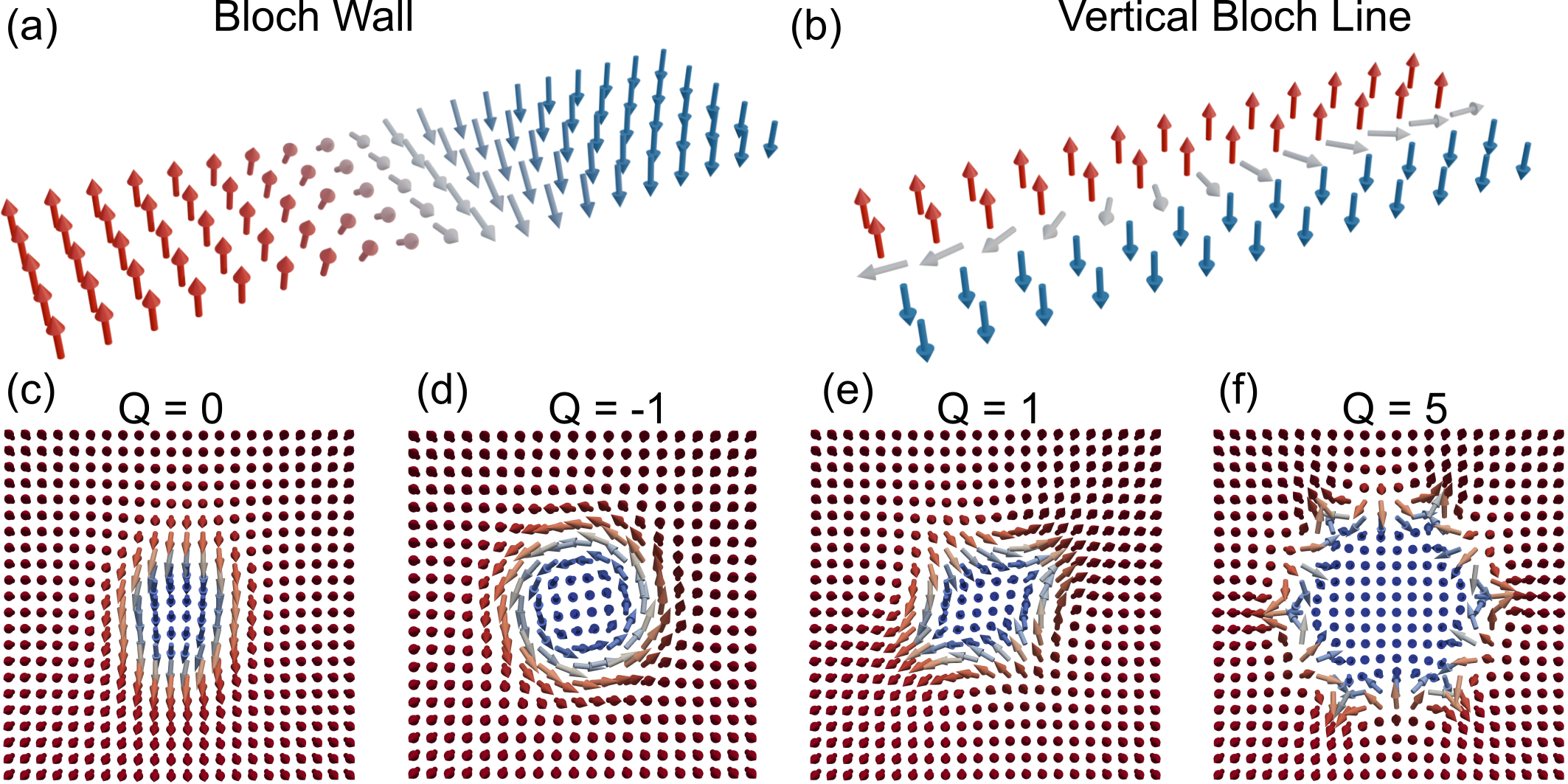}
  \caption{Schematic illustration of the spin configuration of (a) a Bloch wall, (b) a vertical Bloch line, and (c) various spin objects including a trivial bubble with Q = 0, a Bloch skyrmion with Q = -1, an antiskyrmion with Q = 1, and a higher-order antiskyrmion with Q = 5.}
  \label{fig:fig1}
\end{figure*}

Non-uniform magnetization configurations have been envisioned to be useful in a wide variety of technological applications, ranging from magnetic vortices for magnetic field sensors~\cite{suess2018topologically, leitao2024enhanced} to domain wall-based race track memories~\cite{parkin2008magnetic} or logical devices~\cite{allwood2005magnetic, luo2020current, kumar2022domain}. Considering the non-uniform one-dimensional textures of the domain walls, they were desirable for spintronic devices because of their ability to control them with external fields and electrical currents~\cite{catalan2012domain}. Later, two-dimensional spin textures known as magnetic skyrmions gained popularity~\cite{roessler2006spontaneous, yu2010real}. Skyrmions are topologically nontrivial spin objects. Their annihilation into a trivial magnetic state (e.g., a uniformly magnetized state) often requires overcoming a significant energy barrier. This barrier might arise because changing the topological winding number $Q$ could involve singularities (discontinuities) in the magnetization field, which are energetically costly. Because the skyrmions can be stable under external magnetic fields at ambient temperature in specific materials, and they can be set into motion by electric currents and temperature gradients, they have been proposed to be used in skyrmion racetrack memory devices~\cite{fert2013skyrmions, tomasello2014strategy, fert2017magnetic} or for reservoir and unconventional computing tasks~\cite{pinna2020reservoir, raab2022brownian, lee2024task, beneke2024gesture}. With the discovery of their topological counterparts, the antiskyrmions, improved schemes for skyrmion and antiskyrmion-based memory devices have been proposed~\cite{hoffmann2021skyrmion}. A key challenge here is to find a suitable material in which both objects can co-exist. So far, this has only been observed in a handful of materials such as non-centrosymmetric Heussler compounds~\cite{nayak2017magnetic, peng2020controlled, jena2020elliptical}, chiral bulk magnet FeGe~\cite{zheng2022skyrmion}, and ferrimagnetic multilayers~\cite{heigl2021dipolar}. For applications such as reservoir computing where the skyrmionic materials act as the reservoir fabric and the spin objects are excited by high-frequency alternating currents or fields, it would be beneficial to use a high number of distinct spin objects with similar properties to increase the level of nonlinearity in the system. It was shown that materials exist where skyrmions with topological numbers ($Q$) other than one can be stabilized in the form of skyrmion bags and bundles using chiral magnets, which are stable at low temperatures~\cite{foster2019two, tang2021magnetic, yang2024embedded} and room temperature~\cite{zhang2024stable}. Recently, it was experimentally demonstrated that skyrmions and antiskyrmions with arbitrary topological charge are intrinsically stable in ferromagnetic Co/Ni multilayers~\cite{hassan2024dipolar}. The near-compensation of the positive perpendicular magnetic anisotropy and the negative shape anisotropy energy densities allows the formation of 0-dimensional topological defects inside regular Bloch-type domain walls (Fig.~\ref{fig:fig1}a), known as vertical Bloch lines (vBL). The spin configuration of a vBL is depicted in Fig.~\ref{fig:fig1}b. When subjected to an out-of-plane (OOP) magnetic field, the domain walls shrink and form trivial bubbles (Fig.~\ref{fig:fig1}c), skyrmions (Fig.~\ref{fig:fig1}d) or antiskyrmions (Fig.~\ref{fig:fig1}e), as well as higher-order skyrmions and antiskyrmions. A Q = 5 antiskyrmion is depicted in Fig.~\ref{fig:fig1}f. So far, they have been only observed using Lorentz transmission electron microscopy (LTEM), which requires the multilayer to be deposited onto an electron-transparent Si$_3$N$_4$ membrane. Samples thus have to be specially prepared for a successive LTEM study, and the preparation on a thin membrane limits heat dissipation making it challenging to study devices with imposed currents.

In this Letter, we show that first-order antiskyrmions, as well as higher-order spin textures, present in Co/Ni multilayers can be mapped by high-resolution magnetic force microscopy. The presence of antiskyrmions and vertical Bloch lines is confirmed by LTEM measurements using the same film deposited on a $\SI{30}{nm}$ ${\rm Si_3N_4}$ membrane. Furthermore, we calculate the MFM contrast from the magnetization states obtained by micromagnetic simulations permitting the identification and discrimination of the various spin objects observed in high-resolution MFM imaging.

\begin{figure}
  \centering
  \includegraphics[width=\columnwidth]{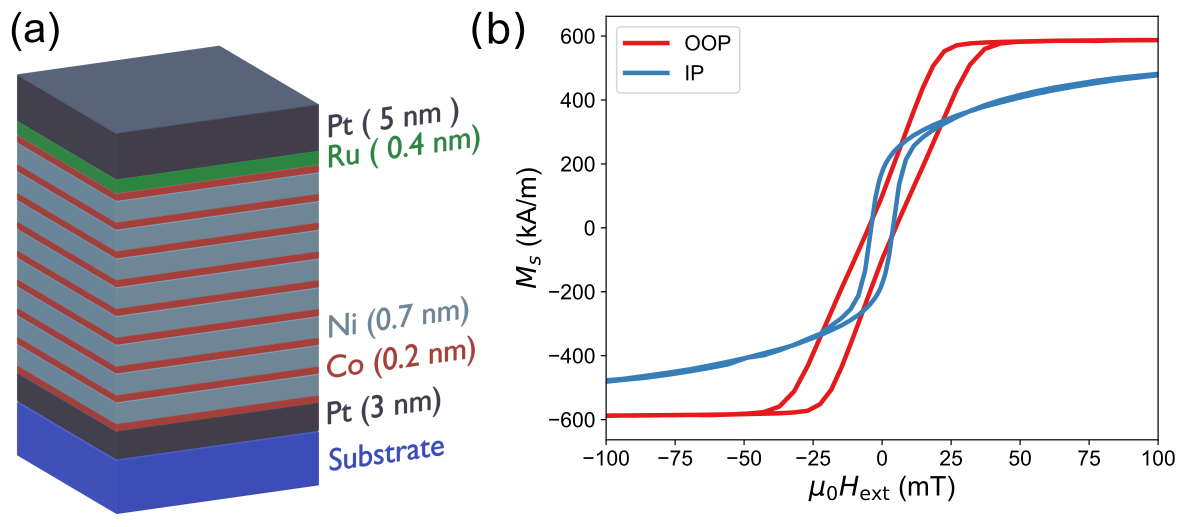}
  \caption{(a) Schematic illustration of the investigated magnetic multilayer stack: Si-SiOx substrate / Pt(3)/[Co(0.2)Ni(0.7)]$_{\times 8}$/Co(0.2)/Ru(0.4)/Pt(5) (thicknesses in nanometer). (b) $M(H)$ hysteresis loops of the sample measured for both OOP (red) and IP (blue) directions.}
  \label{fig:fig2}
\end{figure}

A Co/Ni multilayer thin film was deposited via magnetron sputtering at room temperature on a thermally oxidized Si(100) substrate. The base pressure inside the ultrahigh vacuum chamber was below $\SI{1e-8}{mbar}$. The materials were DC magnetron sputtered from elemental targets utilizing Ar gas at a sputter pressure of $\SI{3.5}{mbar}$. A schematic illustration of the deposited layer stack is given in Fig.~\ref{fig:fig2}a. The total stack is Si-SiOx substrate / Pt(3)/[Co(0.2)Ni(0.7)]$_{\times 8}$/Co(0.2)/Ru(0.4)/Pt(5) with thicknesses given in nanometers. Within the same deposition run, we also included a $\SI{30}{nm}$ SiN membrane to be used for LTEM measurements. Magnetic hysteresis loops were measured for both OOP and in-plane (IP) fields with a superconducting quantum interference device vibration sample magnetometer (SQUID-VSM). The $M(H)$ loops are presented in Fig.~\ref{fig:fig2}b. The OOP loop is typical for a system with perpendicular magnetic anisotropy, which is, however, too weak to keep the film in the saturated state. Hence, domains form when the field is removed. The IP loop shows hysteretic behavior for fields below $\SI{25}{mT}$, and a non-hysteretic increase in magnetization, reaching saturation only at larger fields. We attribute this non-hysteretic part to a rotation of the magnetization into the plane for large IP fields. The hysteresis that occurs for smaller fields arises from the magnetic moments inside the domain walls that align parallel to the applied field in the plane~\cite{Marioni2006}. 

From the same IP hysteresis curve we approximate the effective anisotropy field $\mu_0H_k^{\mathrm{eff}} = \SI{118}{mT}$. With a saturation magnetization $M_\mathrm{s} = \SI{550}{kA/m}$, the perpendicular magnetic anisotropy can be calculated as $K_\mathrm{u} = K_\mathrm{eff} + \frac{1}{2}\mu_0M_s^2$, where $K_{\mathrm{eff}} = \SI{32.5}{kJ/m³}$ from $K_\mathrm{eff} = \frac{1}{2}\mu_0H_k^{\mathrm{eff}}M_\mathrm{s}$. Thus, we obtain $K_\mathrm{u} = \SI{222.4}{kJ/m^3}$. When comparing these values with the stability phase diagram of higher order spin textures given in Ref.~\cite{hassan2024dipolar}, we see that our values fit inside the stability region, but in the lower left corner, where mostly antiskyrmions and very few higher order spin textures should be stable.

\begin{figure*}
  \centering
  \includegraphics[width=\textwidth]{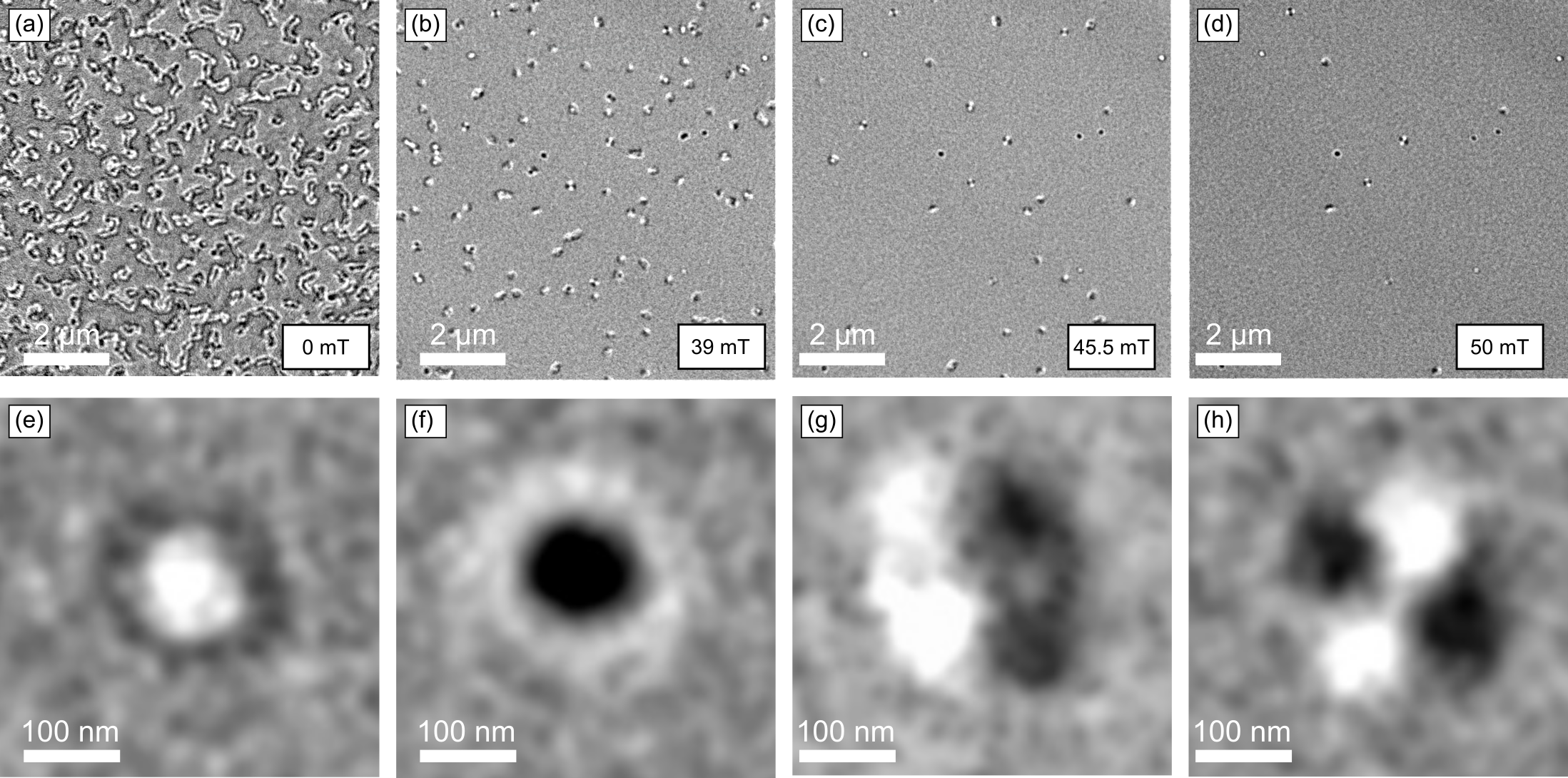}
  \caption{LTEM micrographs taken at room temperature at an external OOP field of (a) $\SI{0}{mT}$, (b) $\SI{30}{mT}$, (c) $\SI{45.5}{mT}$, and (d) $\SI{50}{mT}$. Zoomed in images of an isolated (e) clockwise and (f) counterclockwise skyrmion, (g) a trivial bubble, and (h) an antiskyrmion.}
  \label{fig:fig3}
\end{figure*}

To confirm that spin textures are present in our samples, we use LTEM in Fresnel mode. The LTEM contrasts are acquired at the $\SI{-2}{mm}$ defocus value to visualize the IP components of the magnetization. The background subtracted LTEM results are shown in Fig.~\ref{fig:fig3}, where we followed the approach from Ref.~\cite{denneulin2021visibility} and use a high band pass filter to enhance visibility of spin-objects. Fig.~\ref{fig:fig3}(a) shows the zero-field multidomain states which typically exhibits many vBLs in the domain walls. The evolution of the magnetization state with the applied OOP field is shown in Figs.~\ref{fig:fig3}(b) to (d). With an increasing field, the domains with magnetization antiparallel to the field shrink and collapse into various types of spin texture. 
Figure~\ref{fig:fig3}(e) and (f) show zoomed-in images of skyrmions with clockwise and counterclockwise circularities, respectively. Additionally, many trivial bubbles can be observed as examplified in Fig.~\ref{fig:fig3}(g). Finally, a first-order antiskyrmion is shown in Fig.~\ref{fig:fig3}(h).

\begin{figure*}
  \centering
  \includegraphics[width=\textwidth]{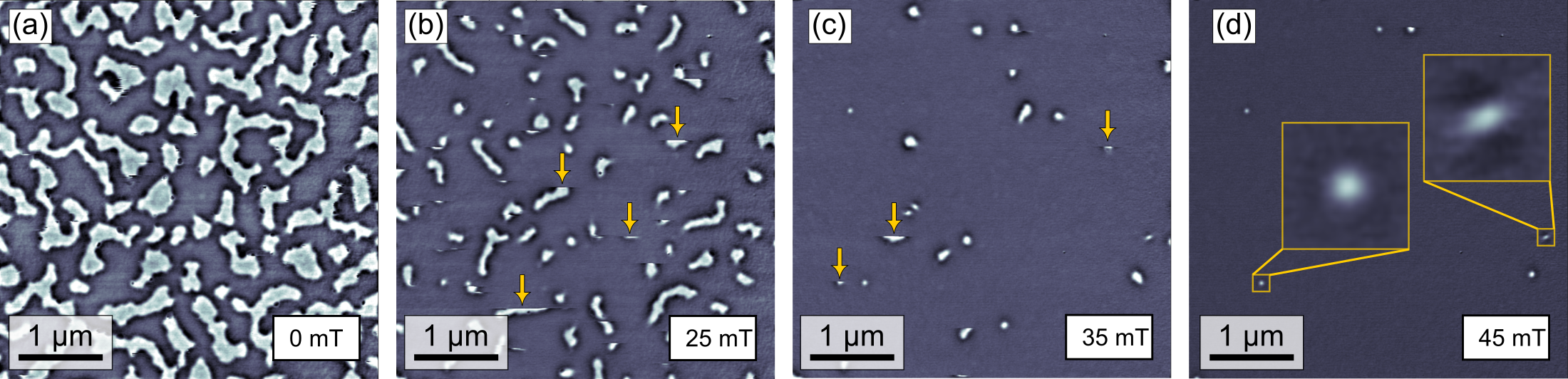}
  \caption{MFM data acquired with both the background measured in saturation and distortion arising from the cantilever canting removed. Applied OOP fields were (a) 0 mT, (b) 25 mT, (c) 35 mT, and (d) 45 mT. A highpass filter is applied to the images to better highlight the frequency shift originating from the local stray fields. The yellow arrows indicate scan lines where a sudden relaxation of the micromagnetic structure during scanning was observed. The yellow squares with a size of $200\times \SI{200}{nm^2}$ in (d) highlight zoomed images of the observed spin textures showing a skyrmion and an antiskyrmion.}
  \label{fig:fig4}
\end{figure*}

In contrast to LTEM, MFM can be performed on magnetic thin film sample deposited on any thick substrate. Here we use a home-built high-resolution magnetic force microscope (MFM) operated under vacuum conditions. To maximize the signal-to-noise ratio and spatial resolution, an uncoated Team Nanotec SS-ISC cantilever with a tip radius smaller than \SI{5}{nm}, a (measured) resonance frequency of $\SI{55,859}{kHz}$ and a force constant of $\SI{1.16}{N/m}$ calculated from the cantilever length $(\SI{225}{\mu m})$, width $(\SI{35}{\mu m})$ and resonance frequency is employed. 
The tip was made sensitive to magnetic stray fields by coating a Ta($\SI{2}{nm}$)/Co($\SI{4.5}{nm}$)/Ta($\SI{3}{nm}$) film on the top side of the tip facing towards the cantilever chip. Magnetic layer deposition was performed with a mask to avoid coating the cantilever towards its attachment point to the chip to obtain a high quality factor $Q=\SI{150e3}{}$~\cite{Feng2022}. The cantilever is then driven on resonance by a phase-looked loop at an oscillation amplitude $A = \SI{5}{nm}$. A single-passage scan mode is used where the tip-sample distance is controlled by a frequency-modulated capacitive feedback as described in our earlier work~\cite{Zhao2018}. With this operation mode, the tip never makes contact with the sample, and hence does not wear. A tip sample distance of $\SI{10}{nm}$ is kept constant on average and does not follow the local topography, with a precision of more than \SI{0.5}{nm} for many days and for different applied magnetic fields. This permits differential imaging techniques, for example, to subtract a background, i.e. the MFM data obtained in saturation from data measured at different fields, which in return allows to eliminate the potential topographic contrasts.´

\begin{figure*}
  \centering
  \includegraphics[width=\textwidth]{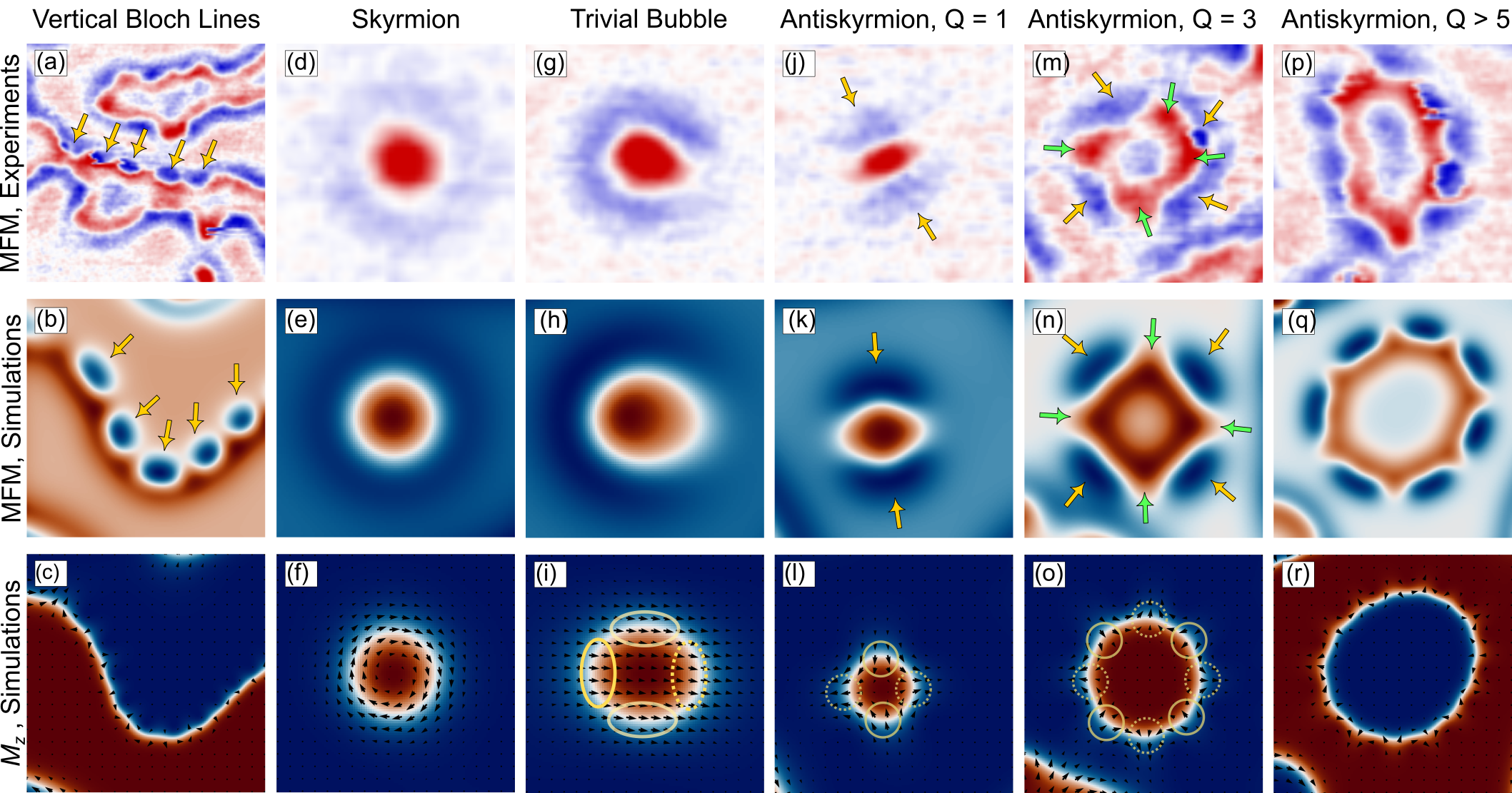}
  \caption{Experimental MFM data (upper row) of different spin objects and the corresponding simulated MFM contrasts at 50 nm height (middle row) and the $M_z$ component of the magnetization (lower row). Following objects are depicted: (a-c) vertical Bloch lines , (d-f) Bloch-type skyrmion, (g-i) trivial bubble, (j-l) antiskyrmion, (m-o) Q = 3 antiskyrmion, and (p) Q>5, (q,r) Q= 5 antiskyrmion. . The yellow arrows show the vertical Bloch lines (a,b) and the high strayfield contrasts sourcing from the kinks of the domain boundary of higher order spin objects. Furthermore, the regions of magnetization generating the \textit{shadows} of stray field around spin objects are highlighted in solid yellow ellipses and circles, while the regions generating no stray field are indicated in dashed ellipses and circles.}
  \label{fig:fig5}
\end{figure*}

Similarly to LTEM, MFM also picks up a magnetic background contrast arising from local variations of the magnetic moment areal density. Additionally, scanning at constant average height leads to a local variation of the van der Waals force (derivative) arising from the (small) sample topography. Subtraction of MFM data obtained with the sample in a saturated state from data measured at smaller fields removes these background contrast contributions, making the small spin textures of the sample more visible. In addition, the cantilever normal is canted by 12$^{\circ}$ relative to the surface normal (which is typical for all scanning force microscopy sets) and, in our case, aligned along the y axis of the images shown in this work. The oscillation of the cantilever along this canted axis then leads to a noticeable asymmetry of the domain walls along the image y axis~\cite{Feng2022}. Because the MFM maps the stray field and not the magnetization, such domain wall asymmetries may impose challenges when different spin textures need to be distinguished. Thus, for all presented MFM images the background measured in saturation and the domain wall asymmetries arising from the canted oscillation of the cantilever have been removed, together with a part of the distance loss~\cite{Feng2022}. 

Analogously to the LTEM data, the MFM reveals submicrometer domain structures under zero field conditions, characterized by domain walls that exhibit various acute bends and kinks attributable to the presence of vBLs (Fig.~\ref{fig:fig4}(a)). At 25 and $\SI{35}{mT}$ smaller domains, circular and elliptical spin textures form (Fig.~\ref{fig:fig4}(b) and (c)), while at $\SI{45}{mT}$ only a few spin objects with sub-100nm diameter are visible. Note that the horizontal lines visible at some locations (see yellow arrows in panels (a) and (b)) arise from a sudden switching events, and thus, spontaneous evolution of the micromagnetic structure (Barkhausen jump) possibly triggered by the stray field of the scanning MFM tip.
Unlike LTEM data, which directly show a measure of the IP magnetization and can hence easily distinguish between skyrmions (Fig.~\ref{fig:fig3}(e) and (f)), trivial bubbles (Fig.~\ref{fig:fig3}(g)) and antiskyrmions (Fig.~\ref{fig:fig3}(h)), the discrimination of these spin objects from MFM images arising from the stray field (derivative) remains more challenging. This is because the stray field is dominated by the up/down area of the spin texture, whereas the vBLs only lead to a weak, very local stray field modification at the domain wall. To enhance the contrast arising from such variations of the small spatial wavelengths, components of the spin texture, and consequently of the stray-field, high-pass filtering is employed. 

In the following, we isolated special cases that we have observed in our MFM experiments. The high-pass-filtered data, derived from the background-subtracted and processed MFM measurements shown in Fig.~\ref{fig:fig4}, are presented using a red/blue color scale in Fig.~\ref{fig:fig5} (top panel). Hereafter, we aim to understand the MFM signatures of different spin textures. Thus, we have a more detailed look at the MFM data, which is an indication of the localized stray fields, and corroborate them with simulated MFM contrasts and magnetization states obtained from micromagnetic simulations.

Our micromagnetic modeling was performed with \texttt{magnum.np}~\cite{magnumnp}, a GPU-accelerated micromagnetic simulation software. We assumed material parameters as in Ref.~\cite{hassan2024dipolar}, namely, $M_s = \SI{940}{kA/m}$, $K_u = \SI{575}{kJ/m^3}$ and $A = \SI{10}{pJ/m}$, as we were only interested in the contrast comparison of the various spin objects. A simulation box with $1024 \times 1024 \times 1$ was discretized in $\SI{2}{nm} \times \SI{2}{nm} \times \SI{5}{nm}$. The magnetization was relaxed at zero magnetic field from a random state by numerically solving the Landau-Lifshitz-Gilbert (LLG) equation. The magnetic field was then increased in \SI{1}{mT} steps and the LLG was solved for $\SI{10}{ns}$ using a high damping value of $\alpha = 0.1$ to ensure that we were in an energetically stable configuration. From the magnetization states, the MFM contrast (here the phase shift $\Delta \phi$) is calculated using a simple point dipole model for the tip, which, however, increases the sharpness or the content of signals with shorter spatial wavelengths. Because of this, we obtain good qualitative agreement with the high-pass filtered experimental MFM data.

 In Fig.~\ref{fig:fig5}(a) the MFM data of the small domains is depicted, which was acquired at zero field. Typically, the MFM contrast of a domain wall separating a domain up and down highlights the localized stray field at the domain wall location. Thus, a simple blue and red adjacent contrasts are expected. In contrast, we observe in Fig.~\ref{fig:fig5}a that the blue dots are enclosed within the domain wall by white rings. Compared with simulated the MFM contrasts from Fig.~\ref{fig:fig5}b, which is calculated from the magnetization profile shown in Fig.~\ref{fig:fig5}c, it becomes obvious that the blue spots indicated with yellow arrows in (a,b) are vertical Bloch lines.

The data displayed in Fig.~\ref{fig:fig5}(d-f) shows a skyrmion characterized by its perfectly circular symmetry with the dark blue \textit{shadow ring} arising from the return of the magnetic flux. From the comparison to MFM simulations and magnetization configuration we can see that we have a clockwise skyrmion.  In contrast to Néel walls~\footnote{Note that for skyrmions with Néel walls, the additional volume magnetic charge density arising from the divergence of the magnetization field in the Néel walls either amplifies or attenuates the skyrmion stray field, making it possible to distinguish counterclockwise and clockwise Néel skyrmions, respectively, from a quantiative analysis of the stray field~\cite{Marioni2018,Bacani2019}.}, Bloch walls, however, do not generate a magnetic volume charge density, and thus Bloch skyrmions with different wall chiralities generate the same stray field and cannot be distinguished by MFM. In contrast, different Bloch wall chiralities lead to a very different appearance of the skyrmions in LTEM images (compare Figs.~\ref{fig:fig3}(e) and (f)). The spin object displayed in Fig.~\ref{fig:fig5}(g) has a noticeably different appearance with its slightly elliptical shape and a partial U-shaped blue \textit{ shadow ring}. Micromagnetic simulations and MFM contrasts (Fig.~\ref{fig:fig5}(h,i)) allow us to categorize this spin texture as a trivial bubble. The counterclockwise and clockwise Néel-type walls on the left (Fig.~\ref{fig:fig5}(i), yellow ellipses, solid lines) and right-hand side (yellow ellipses, dashed lines) lead to a field amplification and attenuation, respectively, making the stray field at the left domain wall sharper and more pronouced than on the right side. Note that the domain wall type at the top and bottom of the trivial bubble in Fig.~\ref{fig:fig5}(i) (light yellow ellipses) remains Bloch-like, generating a wall contrast between that of the left and right sides.

Figure~\ref{fig:fig5}(j) presents the MFM data depicting an antiskyrmion, as confirmed by the simulations in Fig.~\ref{fig:fig5}(k,l). Here, we observe that the MFM contrast of an antiskyrmion has a different signature regarding its magnetic flux \textit{shadow ring}. Instead of a circular core and continuous ring, the antiskyrmions' contrast is distinguished by an elliptical core and the presence of two blue lobes located at the top and bottom (yellow arrows).
An important observation is that a counter clockwise wall structure amplifies the stray field at the top and bottom of the antiskyrmion spin texture leading to a sharp and pronounced change of the stray field (yellow, solid line ellipses in Fig.~\ref{fig:fig5}(l)), whereas the clockwise wall structure (yellow, dashed line ellipses in Fig.~\ref{fig:fig5}(l)). To the best of our knowledge, this is the first experimental high resolution imaging of an isolated \textit{dipolar-stabilized} antiskyrmion.

Panels (m-o) and (p-r) further show higher-order antiskyrmions with topological charge $Q=3$ and possibly $Q \geq 5$. We point out that with increasing topological charge, the radius of the (anti)skyrmions will also increase. Thus, the core becomes more uniformly magnetized compared to the domain wall boundaries. Thus, the resulting stray field and the corresponding MFM contrast is expected to weakenin the core itself, as observed experimentally. In panel (m,n), the contrast in the domain wall is particularly pronounced, with the four dark red lobes marked by green arrows and blue lobes marked by yellow arrows. This is because of stray field amplification from the four counterclockwise Néel wall parts of the domain wall (four yellow, solid line ellipses in Fig.~\ref{fig:fig5}(o)). By quantifying the number of red and blue lobes, we can identify the spin object in panel (m) as $Q=3$. In contrast, the texture observed in panel (p) presents greater challenges in assessment and classification with a similar degree of certainty. Compared with the simulated MFM contrast of an antiskyrmion $Q=5$, it is possible that the texture in (p) has a higher charge than $Q = 5$.

In conclusion, we demonstrated that skyrmions, trivial bubbles, and antiskyrmions of different topological charges present in Co/Ni multilayers can be imaged and distinguished by high-resolution MFM. MFM does not directly image the local magnetization texture, but the stray field (derivative) generated by divergences (magnetic charges) of the magnetization. While the positive and negative magnetic surface charges appearing at the top and bottom surfaces are the dominant sources for the stray field, the magnetic volume charges arising from local Néel wall type spin textures lead to modulations of the wall contrast that can be made visible by high-resolution MFM and permit distinguishing between different types of spin textures. Our results demonstrate that high-resolution MFM is a viable technique to image spin textures with different topological charges which permits the study various spin objects in future devices. 

\textbf{Notes}
The authors declare no competing financial interest.
\subsection*{Acknowledgements}
 S.K. and D.S. acknowledge funding from the Austrian Science Fund (FWF) under grant no. I6267 (CHIRALSPIN). S.K. thanks the Vienna Doctoral School in Physics for funding the Mobility Fellowship. S.K. acknowledges funding from the European Research Council (ERC) under the European Union’s Horizon 2020 research and innovation programme, grant agreement no. 101001290 (3DNANOMAG). We acknowledge Vienna Scientific Cluster for awarding this project access to the LEONARDO supercomputer, owned by the EuroHPC Joint Undertaking, hosted by CINECA (Italy) and the LEONARDO consortium. T.K. and F.B. acknowledge funding by the Helmholtz Young Investigator Group Program through project VH-NG-1520 and by the Deutsche Forschungsgemeinschaft
(DFG, German Research Foundation) through projects 462676630 (BU 3297/3-1) and 49254781 (TRR-360, sub-project C02). M.A. gratefully acknowledges funding from Deutsche Forschungsgemeinschaft (DFG, German Research Foundation) grant no. 507821284. J.S. acknowledges financial support and a scholarship received from the Bavarian Research Foundation (DOK-193-22).

\bibliography{main}

\end{document}